\documentclass{emulateapj}

\shorttitle{Substellar-mass companions to three giant stars}
\shortauthors{Gettel et al.}

\begin{document}

\title{Substellar-Mass Companions to the K-Giants HD 240237, BD +48 738 and HD 96127}

\author{S. Gettel\altaffilmark{1,2}, A. Wolszczan\altaffilmark{1,2}, A. Niedzielski\altaffilmark{3}, G. Nowak\altaffilmark{3}, M. Adam\'ow\altaffilmark{3}, P. Zieli\'nski\altaffilmark{3} \& G. Maciejewski\altaffilmark{3}}

\altaffiltext{1}{Department of Astronomy and Astrophysics, the Pennsylvania State University, 525 Davey Laboratory, University Park, PA 16802, sgettel@astro.psu.edu, alex@astro.psu.edu}

\altaffiltext{2}{Center for Exoplanets and Habitable Worlds, the Pennsylvania State University, 525 Davey Laboratory, University Park, PA 16802}

\altaffiltext{3}{Toru\'n Center for Astronomy, Nicolaus Copernicus University, ul. Gagarina  11, 87-100 Toru\'n, Poland, Andrzej.Niedzielski@astri.uni.torun.pl}

\slugcomment{Oct 21, 2011}

\begin{abstract}
We present the discovery of substellar-mass companions to three giant stars by the 
ongoing Penn State-Toru\'n Planet Search (PTPS) conducted with the 9.2 m Hobby-Eberly 
Telescope. The most massive of the three stars, K2-giant HD 240237, has a 5.3 $M_{J}$ minimum mass companion orbiting the star at a 746-day period. The K0-giant BD +48 738 is orbited by a $\geq$ 0.91 $M_{J}$  planet which has a period of 393 days and shows a non-linear, long-term radial velocity trend that indicates a presence of another, more distant companion, which may have a substellar mass or be a low-mass star. The K2-giant HD 96127, has a $\geq$ 4.0 $M_{J}$ mass companion in a 647-day orbit around the star. The two K2-giants exhibit a significant RV noise that complicates the detection of low-amplitude, periodic variations in the data. If the noise component of the observed RV variations is due to solar-type oscillations, we show, using all the published data for the substellar companions to giants, that its amplitude is anti-correlated with stellar metallicity.
\end{abstract}


\section{Introduction}

Searches for planets around giant stars offer an effective way
to extend studies of planetary system formation and evolution to
stellar masses substantially larger than 1 $M_{\odot}$ \citep{sat03,sat08a,sat10,hat06,dol09,nie07,nie09a,liu09}. Unlike their progenitors on the main-sequence, the cool atmospheres of evolved stars produce many narrow spectral lines, making a radial velocity (RV) measurement precision of $<$ 10 m s$^{-1}$ possible.

There are now over 30 known planets and brown dwarf - mass companions detected around giants, providing statistics 
that are becoming sufficient to constrain the efficiency of planet formation as a function 
of mass and chemical composition of these evolved stars. For example, there is an ongoing debate as to whether the correlation between stellar metallicity and planet frequency present in main-sequence stars \citep{fis05} is weaker or even entirely absent in giants, possibly due to the dilution of metals as the convection zones deepen during giant evolution \citep{pas08,zie11,qui11}. A recent study of a homogeneous sample of subgiants by \citet{joh10} reveals that such correlation does exist in the case of these stars. 

Another interesting phenomenon is the observed deficiency of short-period planets around K-giants, with the closest one detected so far having an orbital radius of 0.6 AU \citep{nie09a}. This is in contrast to subgiant systems, which include a few known ``hot Jupiters'' \citep{joh10}. Models of the dynamics of planetary systems around evolving giants suggest that the effect may be due to the tidal capture and engulfment of planets  at the initial orbital radii of 0.3-0.6 AU \citep{sat08b,vil09,nor10}.

As a consequence of the observed correlation of the frequency of occurrence of massive planets  with stellar mass \citep{lov07,omi11,joh10}, there has been a growing number of detections of brown dwarf - mass companions to giants \citep{lov07,liu08,sat08a,dol09}. These include discoveries of two-companion systems \citep{nie09b,qui11}, in which the orbiting bodies may be brown dwarfs, or supermassive planets created via the core collapse process \citep{ida04}. Recent studies have also uncovered several systems where a dwarf or subgiant is orbited both by a Jupiter-mass planet and a close brown dwarf \citep[e.g.][]{win10,ben10}. In this paper, we report another intriguing case of a system consisting of a Jupiter-mass planet around the giant BD +48 738, and an outer companion, which may be another star, but the currently available constraints do not exclude a possibility that its mass falls in the brown dwarf range.


The RV variations over a wide range of timescales and amplitudes observed in giant stars can be caused by a variety of phenomena, including orbital dynamics, processes that are intrinsic to the star, such as stellar oscillations or spot activity coupled with stellar rotation, and other effects that may still await identification. For example, as shown by the CoRoT and Kepler photometry of giants, both radial and non-radial oscillations are common in these stars \citep{der09,hek10,bau11,gil11}. In fact, p-mode oscillations have been identified in high-cadence RV measurements of several of them \citep[e.g.][]{bar04,hat07,hat08}. Studies by \citet{hek06} and \citet{hek08} demonstrate that one can make a statistical distinction between the stellar and non-stellar origin of RV variations in giants by correlating them with stellar properties, such as the B-V color and surface gravity. Here, we show two examples of such variability in the K2-giants, HD 240237 and HD 96127. Also, using the empirically established dependencies between the effective temperature, $T_{\rm{eff}}$, and the B-V color \citep{alo99}, and $T_{\rm{eff}}$ and surface gravity, log($g$), \citep{hek07}, we investigate the intrinsic RV variability in a sample of K-giants with planets in terms of stellar metallicity, [Fe/H], and find a suggestive anti-correlation between [Fe/H] and the observed stellar jitter.

This paper is organized as follows. An outline of the 
observing procedure and a description of the 
basic properties of the three stars are given in Section 2, followed by the 
analysis of RV measurements in Section 3. The accompanying analysis
of rotation and stellar activity indicators is given in Section 4.
Finally, our results are summarized and further discussed in
Section 5.

\section{Observations and Stellar Parameters}

Observations were made with the Hobby-Eberly Telescope \citep[HET;][]{ram98} 
equipped with the High-Resolution Spectrograph \citep{tul98}, in the 
queue-scheduled mode \citep{she07}. The spectrograph was used in the 
R = 60,000 resolution mode with a gas cell (I$_{2}$) inserted into the optical 
path, and it was fed with a 2'' fiber. Details of our survey, the observing 
procedure, and data analysis have been described in detail elsewhere 
\citep{nie07,nie08}.

The atmospheric parameters of the three stars were determined as part of an extensive study of the PTPS targets described in the forthcoming paper by \citet[][Z11]{zie11}. Here, we give a brief description of the methodology employed in that work.

Measurements of $T_{\rm{eff}}$, log($g$), and [Fe/H] were based on the spectroscopic method of \citet{tak02,tak05a}, in which the Fe\,I and Fe\,II lines are analyzed under the LTE assumption. We used equivalent width measurements of 190 non-blended, strong Fe\,I and 15 Fe\,II lines per star which resulted in the mean uncertainties of 13 K, 0.05, and 0.07 for $T_{\rm{eff}}$, log($g$), and [Fe/H], respectively. Except for [Fe/H], these uncertainties represent numerical errors intrinsic to the iterative procedures of \citet{tak05a,tak05b}. The actual uncertainties are likely to be 2-3 times larger. Uncertainties of the [Fe/H] measurements were computed as standard deviations of the mean Fe\,I and Fe\,II abundances and are thus more realistic.

Since the Hipparcos parallaxes of our stars are either unavailable or very uncertain, only rough estimates of their masses and radii could be made. Initial estimates of stellar luminosities were derived from the $M_{\rm{V}}$ values resulting from the empirical calibrations of \citet{str81}. The intrinsic color indices, (B-V)$_{0}$, and bolometric corrections, $BC_{\rm{V}}$, were calculated from the empirical calibrations of \citet{alo99}.

Stellar masses and ages were estimated by comparing positions of the three stars in the [log($L/L_{\rm\sun}$), log($g$), log($T_{\rm{eff}}$)] space with the theoretical evolutionary tracks of \citet{gir00} and \citet{sal00} for the closest respective values of metallicity. Stellar mass estimates derived from this modeling process were then used to adopt the final values of luminosity. Stellar radii and the associated uncertainties were then computed using the spectroscopic values of $T_{\rm{eff}}$ and adopted luminosities.

Obviously, uncertainties affecting the position of a star in the parameter space determine the corresponding errors in mass estimates. As the values of T$_{\rm{eff}}$ and log($g$) are relatively well determined, the largest uncertainties are contributed by stellar luminosity or parallax. As a useful guidance, the mean uncertainty in stellar mass estimates for over 300 Red Clump giants analyzed in Z11 using the above approach amounts to about 0.3 M$_{\rm\sun}$.


The stellar rotation velocities were estimated by means of the \citet{ben84} cross-correlation method. The cross-correlation functions were computed as described by \citet{now10} with the template profiles cleaned of the blended lines. Given the estimates of stellar radii, the derived rotation velocity limits were used to approximate the stellar rotation periods. The parameters of the three stars are summarized in Table \ref{table1}.

\section{Measurements and Modeling of Radial Velocity Variations}

RVs were measured using the standard I$_{2}$ cell calibration
technique \citep{but96}. A template spectrum was constructed
from a high-resolution Fourier transform spectrometer
(FTS) I$_{2}$ spectrum and a high signal-to-noise stellar spectrum
measured without the I$_{2}$ cell. Doppler shifts were derived from
least-squares fits of template spectra to stellar spectra with the
imprinted I$_{2}$ absorption lines. Typical SNR for each epoch was $\sim$200, as measured at the peak of the blaze function at 5936 \AA. The RV for each epoch was derived
as a mean value of 391 independent measurements from
the 17 usable echelle orders, each divided into 23, 4-5 \AA\ blocks, with a typical, intrinsic uncertainty
of 6-10 m s$^{-1}$ at 1$\sigma$ level over all blocks \citep[][in preparation]{now11}. This RV precision levels made it
quite sufficient to use the \citet{stu80} algorithm to refer the
measured RVs to the Solar System barycenter.

\begin{figure}
\centering
\includegraphics[width=0.5\textwidth,angle=270]{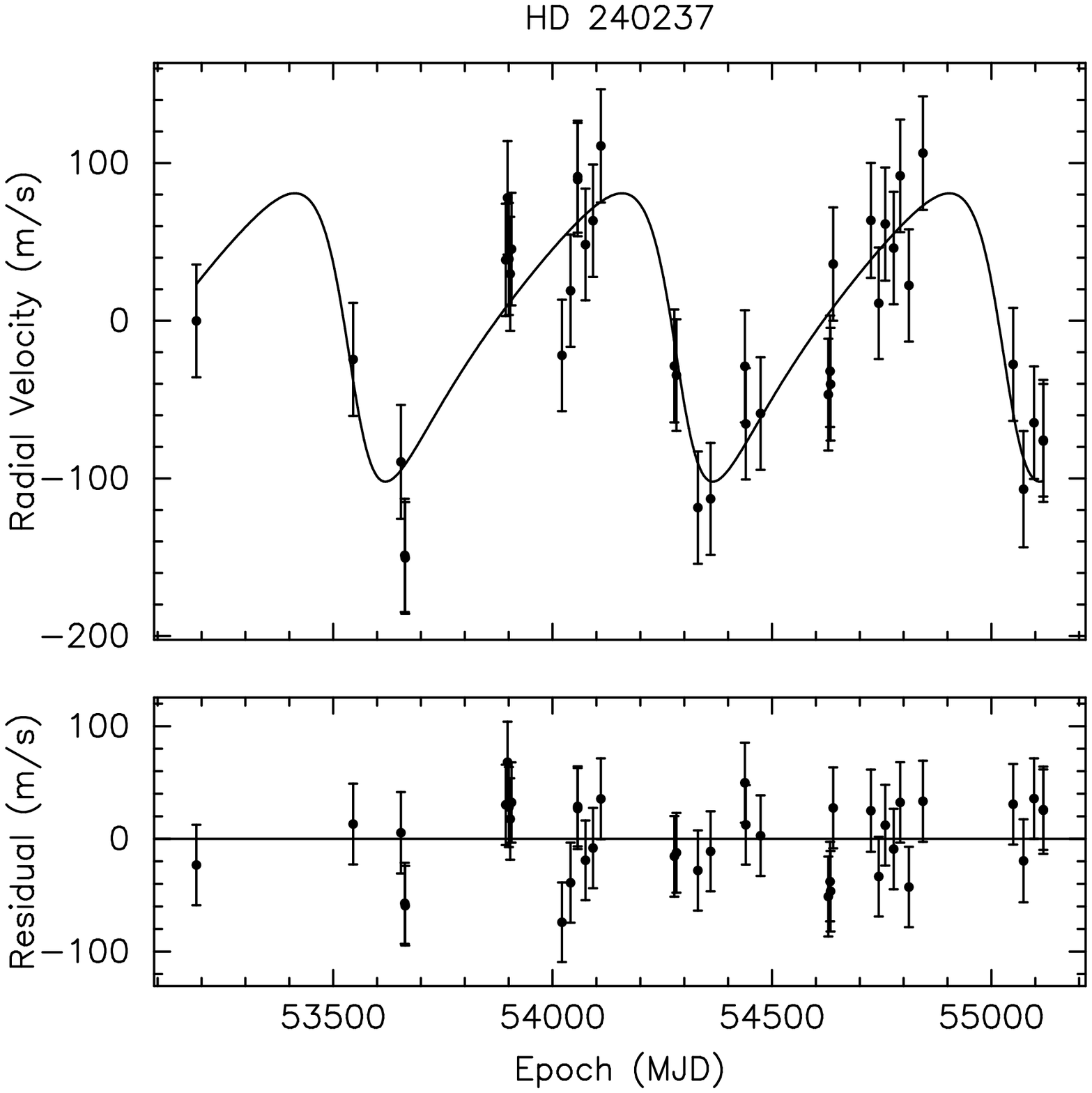}
\caption{Top: Radial velocity measurements of HD 240237 (circles) and the best fit single planet Keplerian model (solid line). Bottom: The post-fit residuals for the single planet model.  \label{fig1}}
\end{figure}

\begin{figure}
\centering
\includegraphics[width=0.5\textwidth,angle=270]{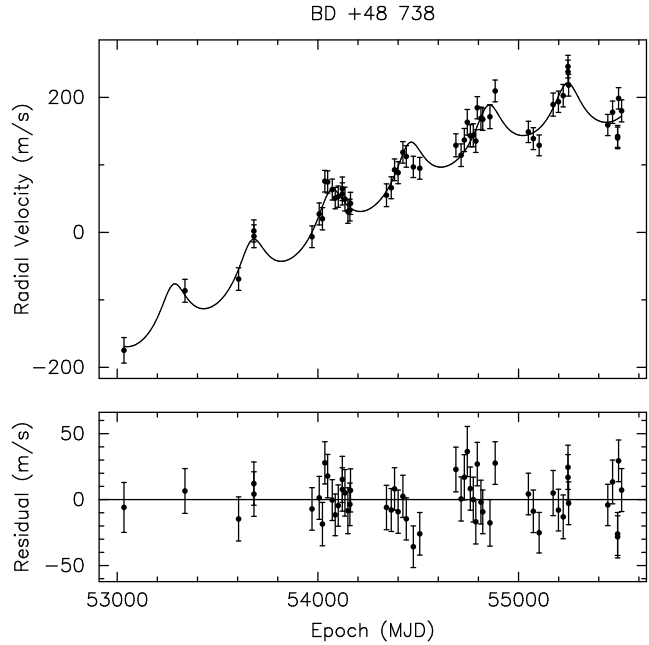}
\caption{Top: Radial velocity measurements of BD +48 738 (circles) and the best fit model consisting of a circular approximation for the long-period orbit, and a 390-day planetary orbit (solid line). Bottom: The post-fit residuals for the above model.
  \label{fig2}}
\end{figure}

The RV measurements of each star were modeled in terms of the standard,  
six-parameter Keplerian orbits, as shown in Figs. \ref{fig1}, \ref{fig2} and \ref{fig4}. Least-squares fits to the data were performed using the Levenberg-Marquardt algorithm \citep{pre92}. Errors in the best-fit orbital parameters were estimated from the parameter covariance matrix. These parameters are listed in Table \ref{table2} for each of the three stars. The estimated, $\sim$7 m s$^{-1}$ errors in RV measurements were evidently too small to account for the actually measured post-fit rms residuals (Table \ref{table2}). As outlined in Section 1, giant stars are subject to atmospheric fluctuations and radial and non-radial pulsations that may manifest themselves as an excess RV variability. In particular, the p-mode oscillations, typically occurring on the timescales of hours to days, are usually heavily under-sampled in the RV surveys of giants and can account for a significant fraction of the observed post-fit RV noise \citep{hek06}. A rough estimate of such variations can be made using the scaling relation of \citet{kje95}, which relates the amplitude of p-mode oscillations to the mass, $M$, and the luminosity, $L$, of the star:
\begin{equation}
v_{osc} = \frac{L/L_{\odot}}{M/M_{\odot}}(23.4 \pm 1.4) 
\end{equation}
where the amplitude, $v_{osc}$ is in cm s$^{-1}$. 

The statistical significance of each detection was assessed by calculating false alarm probabilities (FAP) using the RV scrambling method \citep[][and references therein]{wri07}. FAPs for HD 240237 and HD 96127 were calculated for the null hypothesis that the planetary signal can be adequately accounted for by noise. Two FAP values were calculated for BD +48 738, first for the null hypothesis that the long-term trend can be treated as linear, then for the null hypothesis that it can be modeled as a fraction of a long-period, circular orbit and noise. 

\subsection{HD 240237}

RVs of HD 240237 (BD +57 2714, HIP 114840) are listed in Table \ref{table3}. They were measured at 40 epochs over a
period of 1930 days from July 2004 to October 2009. The SNR values ranged from 161 to 450. The exposure time was selected according to actual weather conditions and ranged between 184 and 900 s.
The estimated mean RV uncertainty for this star was 
7 m s$^{-1}$. 

Radial velocity variations of HD 240237 over a 5-year period are
shown in Fig. \ref{fig1}, together with the best-fit model of a Keplerian
orbit. The post-fit residuals are characterized by
the rms value of 36 m s$^{-1}$ and the FAP for this signal is $<$ 0.01\%. The observed RV variations point to an orbiting companion, which moves in a 746-day,
 eccentric orbit, with a semi-major axis of 1.9 AU, and
has a minimum mass of $m_{2}$ sin $i$ = 5.3 $M_{J}$ for the assumed stellar
mass of 1.7 M$_{\odot}$. For randomly oriented orbital inclinations, there is a $\sim$90\% probability that the companion is planet-mass.

The post-fit residuals for this model exhibit a large, 36 m s$^{-1}$ rms noise. Assuming that this RV jitter is due to solar-like oscillations, the Kjeldsen-Bedding relation predicts an even larger amplitude of $\sim$45 m s$^{-1}$ for this star. To account for these variations, we have quadratically added 35 m s$^{-1}$ to the RV measurement uncertainties before performing the least-squares fit of the orbit. 

\subsection{BD +48 738}

RVs of BD +48 738 (AG +49 313) were measured at 54 epochs over a
period of 2500 days from 
January 2004 to November 2010 (Table \ref{table4}). 
The exposure times ranged between 420 and 1200 seconds depending on observing conditions. The SNR ratio ranged between 90 to 342.
The estimated median RV uncertainty for this star was 
7 m s$^{-1}$.

The RV measurements of BD +48 738 over a 6-year period are shown in Fig. \ref{fig2}. They are characterized by a long-term upward trend, with the superimposed periodic variations. The best-fit Keplerian orbit model for the periodic component, with a circular orbit approximation for the observed trend, is given in Table \ref{table2}. For the assumed stellar mass of 0.7 $M_{\odot}$, the planetary companion has a minimum mass of $m_{2}$ sin $i$ = 0.91 $M_{J}$, in a 393-day, moderately eccentric orbit with a semi-major axis of 0.95 AU. The FAP for this signal is $<$ 0.01\%. Within errors, this solution is not sensitive to a particular choice of eccentricity to model the outer orbit.

\begin{figure}
\centering
\includegraphics[width=0.5\textwidth,angle=0]{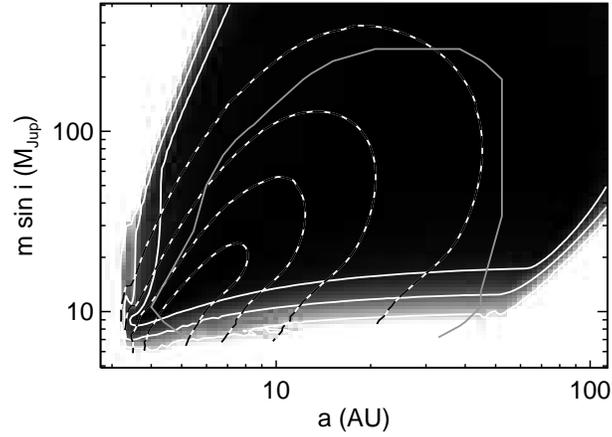}
\caption{Contours of $\chi^{2}$ and $e$ in $a$ - $m$ sin $i$ space of best-fit orbits to the RV data of BD +48 738 with $\chi^{2}$ in gray scale. The solid contours mark the levels at which $\chi^{2}$ increases by 1, 4 and 9 from the minimum, which corresponds to the 1$\sigma$, 2$\sigma$ and 3$\sigma$ confidence intervals. The dashed contours approximately mark the levels of the eccentricity of 0.2, 0.4, 0.6 and 0.8, from the inner to the outer contour, respectively. Orbits interior to the grey contour are found to be stable for at least $10^{4}$ years. \label{fig3}}
\end{figure}

\begin{figure}
\centering
\includegraphics[width=0.5\textwidth,angle=270]{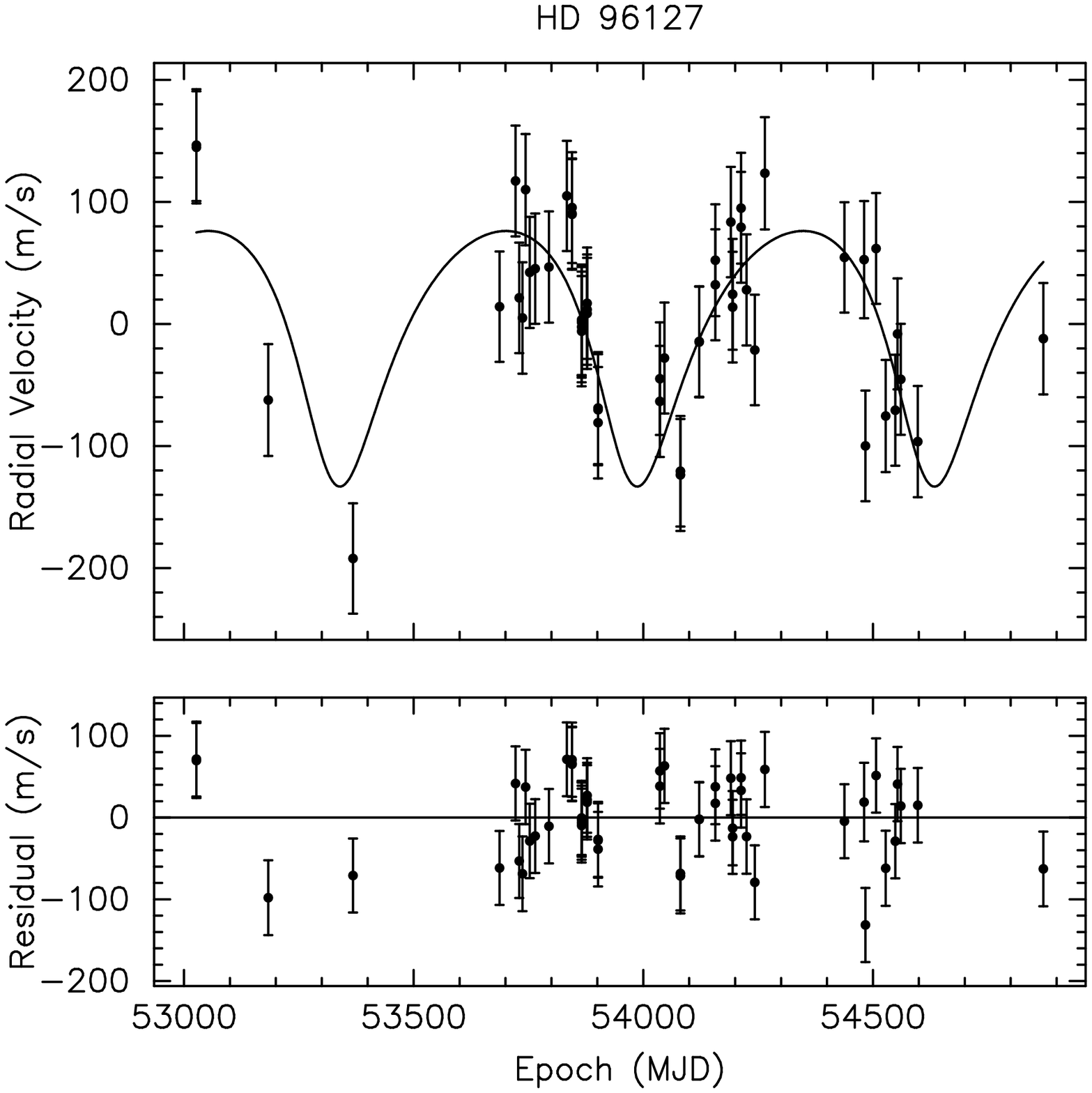}
\caption{Top: Radial velocity measurements of HD 96127 (circles) and the best fit single planet Keplerian model (solid line). Bottom: The post-fit residuals for the single planet model.  \label{fig4}}
\end{figure}

A simultaneous characterization of the two orbits is difficult, as the outer companion's period is much longer than the time baseline of our measurements. However, the curvature of the observed trend is statistically significant, with a FAP value of 0.01\%. This makes it feasible to constrain a range of possible two-orbit models by fitting for parameters of the inner planet over a grid of fixed solutions for the long-period orbit. We have chosen to search for a $\chi^{2}$ minimum over a wide range of values in the $a - m$ sin $i - e$ space following the method of \citet{wri07}. The results are shown in Fig. \ref{fig3} with contours marking the 1, 2 and 3-$\sigma$ confidence levels computed under the assumption that the post-fit residual noise has a Gaussian probability distribution. 

To place limits on a range of the wide orbit models allowed by the available RV data, we considered solutions within the 2-$\sigma$ contour and further restricted them to those with $e <$ 0.8, above which the occurrence of low mass stellar companions to F7-K primaries falls off \citep{hal05}. For the same reason, models involving planet-mass companions were limited to those with $e <$ 0.6 \citep{but06}. We tested the stability of these orbits, by carrying out many N-body simulations using the MERCURY code \citep{cha99}.
Orbits from the grid in Fig. \ref{fig3} were integrated for $10^{4}$ years over a range of $a$, $m$sin $i$ and $e$ values within the 2-$\sigma$ contour.  

As Fig. \ref{fig3} clearly shows, the curvature of the long-period orbit is not yet sufficient to produce tight constraints on the outer companion to BD +48 738. Although only a narrow range of massive planets, out to $a\sim $10 AU, is still allowed, it is much more likely that the companion's minimum mass is in a brown dwarf or a low-mass star range and that its orbit is at least moderately eccentric. Continuing observations of the star will help placing tighter constraints on the orbiting bodies in this interesting system.

\subsection{HD 96127}

RVs of HD 96127 (BD +45 1892) were measured at 50 epochs over a
period of 1840 days from January 2004 to February 2009 (Table \ref{table5}). The SNR for these measurements ranged from 80 to 494. The exposure time was selected according to actual weather conditions and ranged between 72 and 1200 s. The estimated mean RV uncertainty for this star was 6 m s$^{-1}$.

The radial velocity variations of HD 96127 over a 5-year period are
shown in Fig. \ref{fig4}, together with the best-fit model of a Keplerian
orbit. As with HD 240237, the post-fit residuals show a large, 50 m s$^{-1}$ rms RV jitter, comparable to the Kjeldsen-Bedding amplitude estimate of 45 m s$^{-1}$. To account for these variations, we have quadratically added 45 m s$^{-1}$ to the RV measurement uncertainties. The best-fit parameters indicate a planet candidate with a minimum mass of 4 $M_{J}$, and a 647-day, 1.42 AU, moderately eccentric orbit. The FAP for this signal is $<$ 0.01\%.

\section{Stellar Photometry and Line Bisector Analysis}

Photometry and line bisector analysis, which are efficient stellar activity indicators, have been routinely used to verify the authenticity of planet detections by means of the RV method \citep[e.g.][]{que01}. In particular, when contemporaneous photometric data are not available, it is important to correlate the time variability of line bisectors and RVs, in order to investigate a possible contribution of stellar activity to the observed RV behavior. For example, even for slow rotators like K-giants discussed here, a stellar spot only a few percent in size could introduce line profile variations as large as $\sim$100 m s$^{-1}$ \citep{hat02}, which would be easy to detect.

Although a correlation between line bisector and RVs does provide evidence that the intrinsic stellar processes contribute part or all of the observed RV variability, many sources of jitter cannot be readily identified in a bisector analysis. This is because the bisector changes may be too small to measure distinctly from the apparent RV shift or they are indistinguishable from temporal changes in the spectrograph line spread function. Consequently, the lack of such a correlation is a necessary but not a sufficient condition for demonstrating that measured RV variations are of a planetary origin. Additional information, such as a distinctly Keplerian shape of the RV curve \citep{fri02,zec08} or a strict persistence of the RV periodicity over many cycles \citep{qui11}, is necessary to fully verify such detections. 

\begin{figure}
\centering
\includegraphics[width=0.5\textwidth,angle=270]{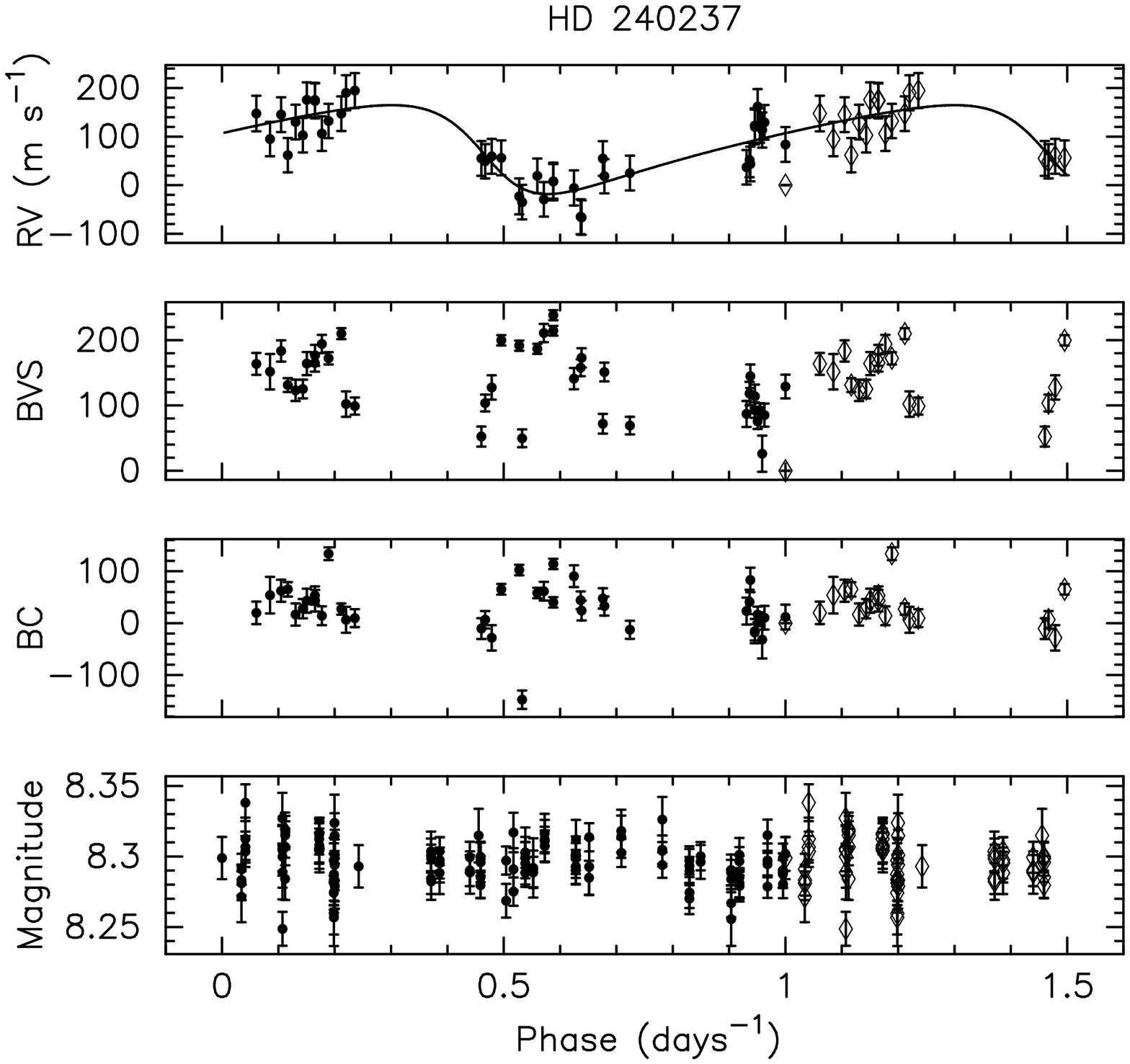}
\caption{HD 240237b data folded at the best-fit orbital period. From top to bottom: (a) Radial velocity measurements with the best fit model (b) Bisector Velocity Span (m/s) (c) Bisector Curvature (m/s) 
(d) Hipparcos photometry.\label{fig5}}
\end{figure}

\begin{figure}
\centering
\includegraphics[width=0.5\textwidth,angle=270]{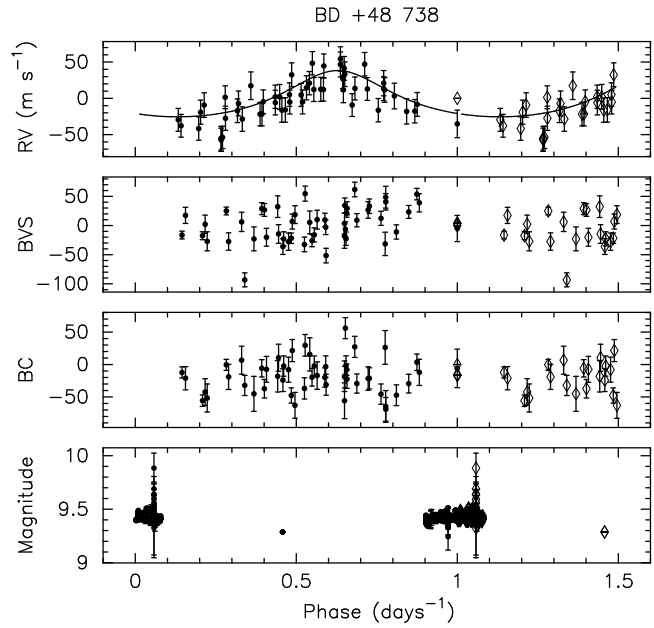}
\caption{BD +48 738b data folded at the best-fit orbital period. From top to bottom: (a) Radial velocity measurements with the best fit inner planet model and the outer orbit removed (b) Bisector Velocity Span (m/s) (c) Bisector Curvature (m/s) (d) WASP photometry.\label{fig6}}
\end{figure}

\begin{figure}
\centering
\includegraphics[width=0.5\textwidth,angle=270]{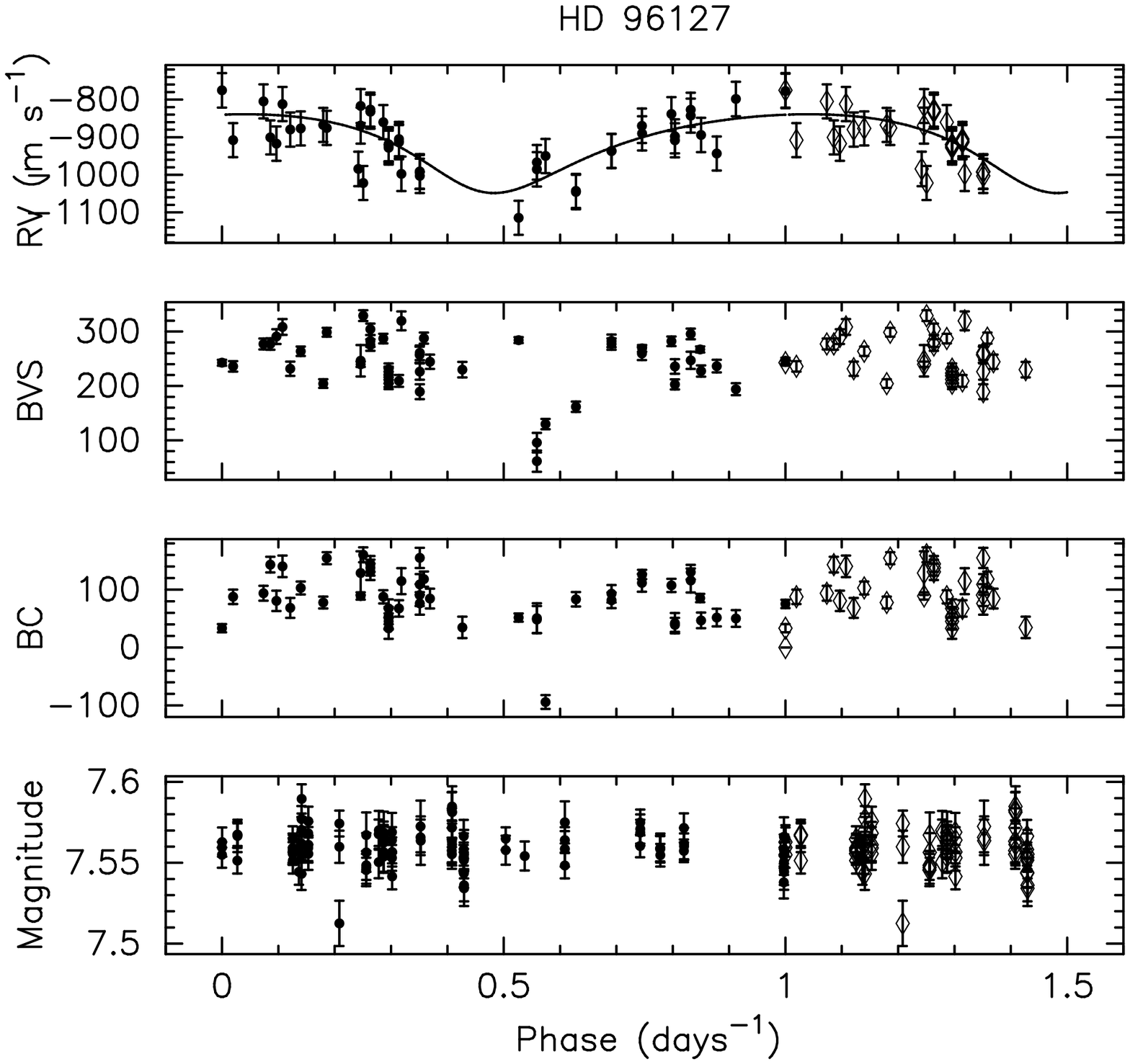}
\caption{HD 96127b data folded at the best-fit orbital period. From top to bottom: (a) Radial velocity measurements with the best fit model (b) Bisector Velocity Span (m/s) (c) Bisector Curvature (m/s) 
(d) Hipparcos photometry.\label{fig7}}
\end{figure}

In order to investigate a possible contribution of stellar jitter to the observed RV periodicities, we have examined the existing photometry data in search for any periodic light variations, and performed a thorough analysis of time variations in line bisector velocity span (BVS) and bisector curvature (BC) for each of the three stars discussed in this  paper. The BVSs and BCs were measured using the cross-correlation method proposed by \citet{mar05} and applied to our data as described in \citet{now10}. For each star and each spectrum used to measure RVs at all the observing epochs, cross-correlation functions were  computed from $\sim$1000 line  profiles with the I$_{2}$ lines removed from  the spectra. The time series for these parameters and the photometric data folded at the observed RV periods for the three stars are shown in Figs. \ref{fig5}, \ref{fig6} and \ref{fig7}.

In addition, using the  scatter seen in  the  photometry  data  and the
rotational velocities determined as described above, we have estimated the amplitude of RV and BVS variations due to a possible presence of spots on the stellar surface \citep{hat02}.  The calculated values range from $<$ 5 m s$^{-1}$ to $<$ 13 m s$^{-1}$ and from $<$ 17 m$^{-1}$ to $<$ 27 m s$^{-1}$, for the RV and the BVS amplitudes, respectively, which is much less than the variability observed in the three stars.

\subsection{HD 240237}

The longest time span photometry available for this star
comes from 120 Hipparcos  measurements \citep{per97}, made between
MJD 47884 and 49042. These data give a mean magnitude of the star of $V = 8.30 \pm
0.02$. There  are also 15  epochs of photometric observations  of this
star available from the Northern Sky Variability Survey \citep{woz04}. Neither of these data sets exhibits periodic brightness variations. 

The mean  values of the BVS  and the BC  for this star are  $136.6 \pm
51.4$ and $29.9 \pm 47.3$m s$^{-1}$. No correlations between the RV variations and those of the BVS and BC were found with the respective correlation coefficients of
r = -0.15  for the BVS  and r = -0.06 for  the BC.   

\subsection{BD +48 738} 

The only photometric database available for this star originates from the WASP measurements \citep{pol06}. The WASP data are contemporaneous with our
RV measurements and give a mean value of $V = 9.41  \pm 0.03 $, with no detectable periodic variations. 
 
We have also derived bisectors and line curvatures  for all the RV
measurement epochs, and obtained the mean values of BVS = 149.6$\pm$30.2 m s$^{-1}$ and BC = 49.4$\pm$27.4 m s$^{-1}$. Time variations of these parameters are not correlated with the observed RV variability and yield the corresponding correlation coefficients of r = 0.11 for the BVS and r = -0.08  for the BC time sequences, respectively.  

\subsection{HD 96127}

Because this star is relatively bright, the only photometry available for it is from 100 Hipparcos measurements \citep{per97}, made between MJD 47892 and 48933.
 The mean magnitude of the star calculated from these data is $V=7.43\pm0.01$. In addition, the time series exhibits a marginal (2$\sigma$), 25.2$\pm$0.05-day periodicity with the peak-to-peak amplitude of 0.02 mag. Analysis of the photometric time series divided into shorter sections shows that this oscillation persists over the entire $\sim$3-year span of the data and, despite its low amplitude, it has a FAP $<$ 0.001\%. On the other hand, it is not present in our RV data for this star and it would be hard to envision how could it be related to the observed 647-day RV periodicity. Moreover, the 4.3-day period corresponding to the frequency of maximum power of p-mode oscillations predicted from \citet{kje95} is obviously much shorter than observed. Evidently, additional high-cadence photometric observations of HD 96127 are needed to clarify the nature of this 25-day periodicity.

The mean  values of the BVS  and the BC  for this star are  $243.3 \pm
52.5$ and $85.2 \pm 44.5$ m s$^{-1}$ and the correlations between the RV and the BVS and BC variations are not significant at r =  0.21  for  the BVS  and r =  0.17 for  the BC, respectively. 

\section{Discussion}

In this paper, we report detections of periodic RV variations in three K-giant stars from the list of targets monitored by the PTPS program. The accompanying analyses of the photometric data, as well as of the time variability of line bisectors and bisector curvature indicate that the most likely origin of the observed periodicities is the Keplerian motion of substellar companions to these stars.

The most interesting of the three stars is the case of the K0-giant, BD +48 738, which, in addition to a $\geq$ 0.91 $M_{J}$ planet moving in a $\sim$1 AU, $\sim$400-day orbit around the star, exhibits a long-term RV trend, which indicates the presence of another, more distant companion. The emerging curvature of the long-term RV trend is not yet sufficient to determine whether the object has a substellar mass, or is a low-mass star.

In this context, it is interesting to note that there has been a growing number of detections of substellar companions to giants that have minimum masses in excess of 10 $M_{J}$, making them candidates for either brown dwarfs or supermassive planets \citep{lov07,liu08,sat08a,dol09}. In two cases, double brown dwarf - mass  companions have been discovered suggesting the possibility that they originated in the circumstellar disk, similar to giant exoplanets \citep{nie09b,qui11}. If the outer companion to BD +48 738 proves to be substellar, we may have another interesting case of an inner Jupiter-mass planet and a more distant, brown dwarf - mass body orbiting the same star.

In principle, such a system could form from a sufficiently massive protoplanetary disk by means of the standard core accretion mechanism \citep{ida04}, with the outer companion having more time than the inner one to accumulate a brown dwarf like mass. A more exotic scenario could be envisioned, in which the inner planet forms in the standard manner, while the outer companion arises from a gravitational instability in the circumstellar disk at the time of the star formation \citep[e.g.][]{kra11}. In any case, it is quite clear that this detection, together with the other ones mentioned above, further emphasizes the possibility that a clear distinction between giant planets and brown dwarfs may be difficult to make \citep[see, for example,][]{opp09}.

The low signal-to-noise planet detections around the K2-giants, HD 240237 and HD 96127 reported in this paper further demonstrate that the intrinsic RV noise in giants restricts the utility of the Doppler velocity method, especially in the case of long periods and large orbital radii, for which its sensitivity degrades for dynamical reasons. This effect has been studied by \citet{hek06}, who have shown that the rms RV noise, $\sigma_{RV}$, of K-giants has a median value of $\sim$20 m s$^{-1}$ and it tends to increase toward later spectral types. Similarly, a negative correlation has been found between $\sigma_{RV}$ and the stellar surface gravity \citep{hat98,hek08}.

A short timescale, low amplitude RV noise commonly observed in giants appears to be the manifestation of undersampled solar-like oscillations \citep{hek06,hek08,qui11}. In particular, the recent photometric studies of K-giants by the CoRoT and Kepler missions \citep{bau11,cia11,gil11,hek11} have shown a prominent presence of such oscillations in these stars. More specifically, \citet{kje95} have proposed that the velocity amplitude associated with the p-mode oscillations scales as $v_{osc}\sim L/M$ (Eqn. 1), or equivalently as $T_{\rm{eff}}^4/g$, where $T_{\rm{eff}}$, and $g$ are the effective temperature and the surface gravity of the star. As shown, for example, by \citet{hek06,hek08} and \citet{dol11}, observations are in a general agreement with these scalings.

In order to compare the observed trend in the RV data for K-giants to the \citet{kje95} prediction more quantitatively, we have used the empirical dependencies of B-V color and log($g$) on $T_{\rm{eff}}$ for giants \citep{alo99,hek07} to generate a semi-analytical function that relates $\sigma_{RV}$ to the observed B-V for different stellar metallicities ([Fe/H]). The observed 30 values of $\sigma_{RV}$ have been taken from the published discoveries of substellar companions to K-giants in Table \ref{table6}, as post-fit residuals after the removal of the Keplerian signal from the RV data. This sample selection has been motivated by the work of \citet{hek08}, which shows that amplitudes of RV periodicities in K-giants are generally independent of stellar parameters, whereas the RV noise left over after their removal tends to correlate with the B-V color and anti-correlate with log($g$). Consequently, our approach practically ensures that the remaining short-term noise exceeding the instrumental RV precision is intrinsic to the star, given the observed absence of planets with orbital radii smaller than 0.6 AU around K-giants \citep{joh07,sat08a,nie09a}. Moreover, as the RV precision and sampling patterns characterizing the ongoing GK-giant surveys are quite similar, any selection effects that could influence the RV jitter measurements can be safely ignored.

\begin{figure}
\centering
\includegraphics[width=0.5\textwidth]{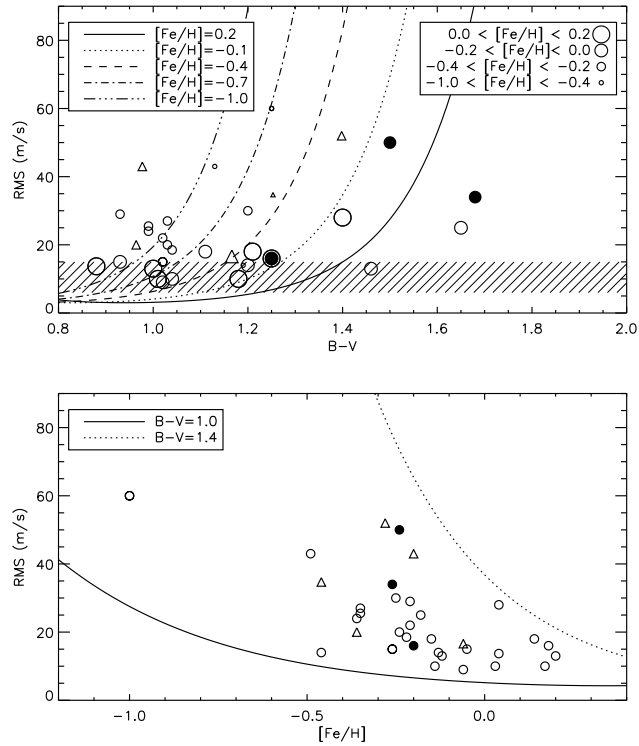}
\caption{Top: Post-fit rms RV error for Keplerian orbit models as a function of the B-V color for giant stars with published substellar companions (open circles), the stars discussed in this paper (filled circles), and stars with new, unpublished planet candidates from this survey (triangles). Theoretical predictions of the p-mode oscillation amplitude as a function of B-V, derived from a modified Kjeldsen-Bedding relationship, are shown as a set of curves for five fixed values of [Fe/H]. The shaded region indicates the range of RV precision of the different giant star planet surveys. Bottom: The same data plotted as a function of the observed [Fe/H]. Also shown are the theoretical curves for two fixed values of B-V. \label{fig8}}
\end{figure}

In Fig. \ref{fig8}, the measured values of $\sigma_{RV}$ as a function of B-V are compared with the Kjeldsen \& Bedding relationship expressed in terms of these two observables as described above. The apparent increase of RV noise with the decreasing stellar metallicity becomes more evident, if plotted out directly. The calculated Pearson correlation coefficient for this relationship is -0.71. The origin of this trend is most likely related to the fact that higher metallicity (opacity) of the star lowers its temperature, which decreases the amplitude of p-mode oscillations, while lower metallicity has the opposite effect. 

One interesting consequence of the observed $\sigma_{RV}$-[Fe/H] dependence in K-giants is that it must affect the analyses of a correlation between the planet frequency and metallicity of these stars \citep{pas07,hek07,ghe10}. No clear agreement has been reached so far, as to whether the observed correlation is weaker than in the case of dwarf stars \citep{fis05}, or not. The effect unraveled here creates an apparent preference for higher metallicity stars to be more frequent hosts to orbiting planets, so that the published studies must be biased by this trend.

We thank the HET resident astronomers and telescope operators for support. SG and AW were supported by the NASA grant NNX09AB36G.  AN and GN were supported in part by the Polish Ministry of Science 
and Higher Education grant N N203 510938 and  N N203 386237. GM acknowledges the financial support from the Polish Ministry 
of Science and Higher Education through the Juventus Plus grant IP2010 
023070.
SG thanks Dr. Jason Wright for many helpful discussions.

The HET is a 
joint project of the University of Texas at Austin, the Pennsylvania State University, Stanford University, Ludwig-Maximilians-Universit\"at M\"unchen, and Georg-August-Universit\"at G\"ottingen. 
The HET is named in honor of its principal benefactors, William 
P. Hobby and Robert E. Eberly. The Center for Exoplanets and 
Habitable Worlds is supported by the Pennsylvania State University, the Eberly College of Science, and the Pennsylvania 
Space Grant Consortium.



\clearpage
\newpage

\begin{deluxetable}{lccc}
\tablewidth{0pt}
\tablecolumns{4}
\tablecaption{Stellar Parameters\label{table1}}
\tablehead{\colhead{Parameter} & \colhead{HD 240237} & \colhead{BD +48 738} & \colhead{HD 96127} }
\startdata
V & 8.19 & 9.14 & 7.43\\
B-V & 1.682 $\pm$ 0.029 & 1.246 $\pm$ 0.046 & 1.503 $\pm$ 0.014\\
Spectral Type & K2 III & K0 III & K2 III\\
$T_{\rm{eff}}$[K]& 4361 $\pm$ 10 & 4414 $\pm$ 15  & 4152 $\pm$23\\
$\pi$[mas]& -- & -- & 1.85 $\pm$ 0.89\\
log($g$) & 1.66 $\pm$ 0.05 & 2.24 $\pm$ 0.06 & 2.06 $\pm$ 0.09\\
$$[Fe/H] & -0.26 $\pm$ 0.07 & -0.20 $\pm$ 0.07 & -0.24 $\pm$ 0.10\\
log($L_{\star}/L_{\odot}$) & 2.49 $\pm$ 0.20 & 1.69 $\pm$  0.33 & 2.86 $\pm$ 0.47\\
$M_{\star}/M_{\odot}$ & 1.69 $\pm$ 0.42 & 0.74 $\pm$ 0.39 & 0.91 $\pm$  0.25\\
$R_{\star}/R_{\odot}$ & 32 $\pm$ 1 & 11 $\pm$ 1 & 35 $\pm$ 17\\
$V_{rot}$[Km/s] & $\le$1 & $\le$1 & $\le$0.5 \\
$P_{rot}$[d] & $\ge$1010 & $\ge$505 & $\ge$1515  \\
\enddata
\end{deluxetable}

\begin{deluxetable}{lccc}
\tablecolumns{4}
\tablewidth{0pt}
\tablecaption{Orbital Parameters\label{table2}}
\tablehead{\colhead{Parameter} &  \colhead{HD 240237b} & \colhead{BD +48 738b} & \colhead{HD 96127b} }
\startdata
P [days] & 745.7$\pm$13.8 &392.6$\pm$5.5 & 647.3$\pm$16.8\\
$T_{0}$ [MJD] & 54292.0$\pm$28.3 & 54457.2$\pm$28.5 & 53969.4$\pm$31.0\\
K [m/s] & 91.5$\pm$12.8 &31.9$\pm$2.6 &104.8$\pm$10.6 \\
e & 0.4$\pm$0.1 &0.2$\pm$0.1 &0.3$\pm$0.1\\
$\omega$ [deg] &108.1$\pm$21.8 &358.9$\pm$31.1 &162.0$\pm$18.2 \\
$m_{2}sin(i)$ [$M_{J}$]&5.3 &0.91  &4.0 \\
$a$ [AU] & 1.9 &1.0 &1.4\\
$\chi^{2}$ &1.1 &1.1 &1.3 \\
Post-fit rms [m/s] & 36.0 & 16.0 & 50.0 \\
\enddata
\end{deluxetable}

\begin{deluxetable}{lcc}
\tablecolumns{3}
\tablewidth{0pt}
\tablecaption{Relative RVs of HD 240237 \label{table3}}
\tablehead{\colhead{Epoch [MJD]} & \colhead{RV [m s$^{-1}$]} & \colhead{$\sigma_{RV}$ [m s$^{-1}$]}}
\startdata
 53188.39793  &     83.7  &      7.3    \\
 53545.43082  &     59.3  &      7.8    \\
 53654.28150  &     -5.7  &      8.9	\\
 53663.09284  &    -65.0  &      8.1	\\
 53664.25192  &    -66.6  &      5.1	\\
 53893.44619  &    122.4  &      6.6	\\
 53897.43174  &    161.8  &      8.3	\\
 53900.44113  &    123.0  &      6.1	\\
 53903.41408  &    113.6  &      8.9	\\
 53906.45211  &    129.3  &      7.0	\\
 54021.25319  &     61.8  &      4.5	\\
 54041.06348  &    102.9  &      6.2	\\
 54057.02965  &    175.2  &      6.0	\\
 54057.16257  &    173.4  &      8.1	\\
 54075.12739  &    132.2  &      5.3	\\
 54092.10329  &    147.2  &      6.7	\\
 54110.05038  &    194.7  &      8.1	\\
 54277.40802  &     55.1  &      7.5	\\
 54282.39529  &     49.3  &      5.3	\\
 54331.26734  &    -34.7  &      6.6	\\
 54360.18436  &    -29.2  &      6.1	\\
 54438.14524  &     54.9  &      6.4	\\
 54440.13141  &     18.4  &      4.7	\\
 54474.05096  &     25.0  &      7.2	\\
 54628.44381  &     37.0  &      5.5	\\
 54632.43929  &     51.8  &      4.7	\\
 54633.42385  &     43.6  &      7.2	\\
 54639.38022  &    119.8  &      8.2	\\
 54725.35386  &    147.5  &     10.3	\\
 54743.31509  &     94.9  &      5.1	\\
 54758.09993  &    145.2  &      7.8	\\
 54777.22166  &    129.9  &      6.8	\\
 54792.16502  &    175.7  &      7.2	\\
 54812.12646  &    106.2  &      6.7	\\
 54844.04772  &    190.2  &      8.4	\\
 55049.45087  &     56.2  &      7.6	\\
 55073.23259  &    -23.0  &     11.5	\\
 55097.18015  &     19.1  &      7.2	\\
 55118.24693  &      7.6  &     16.6	\\
 55118.29165  &      8.1  &      7.1	\\
\enddata
\end{deluxetable}

\begin{deluxetable}{lcc}
\tablecolumns{3}
\tablewidth{0pt}
\tablecaption{Relative RVs of BD +48 738\label{table4}}
\tablehead{\colhead{Epoch [MJD]} & \colhead{RV [m s$^{-1}$]} & \colhead{$\sigma_{RV}$ [m s$^{-1}$]}}
\startdata
53033.65040   &     -721.3  &  11.6  \\
53337.60326   &     -633.0  &   8.0  \\
53604.86380   &     -615.7  &   7.4  \\
53680.90190   &     -544.3  &   6.6  \\
53681.65521   &     -552.3  &   7.8  \\
53970.86815   &     -553.0  &   5.9  \\
54006.77500   &     -519.2  &   6.3  \\
54022.73748   &     -526.3  &   6.7  \\
54034.69078   &     -470.7  &   5.6  \\
54048.65158   &     -471.8  &   6.8  \\
54071.79441   &     -483.0  &   3.8  \\
54086.56153   &     -495.7  &   5.0  \\
54101.71295   &     -493.5  &   4.5  \\
54121.67129   &     -481.9  &   8.9  \\
54121.68284   &     -489.5  &   6.6  \\
54134.63773   &     -497.1  &   9.2  \\
54149.60381   &     -515.6  &   8.8  \\
54159.58385   &     -513.4  &   5.9  \\
54162.57425   &     -503.5  &   6.6  \\
54341.85828   &     -491.4  &   7.6  \\
54365.78790   &     -480.4  &   6.3  \\
54381.96192   &     -453.8  &   5.7  \\
54399.68128   &     -458.0  &   6.6  \\
54423.63082   &     -427.9  &   5.6  \\
54440.79597   &     -433.9  &   5.4  \\
54475.71081   &     -449.5  &   4.8  \\
54507.61437   &     -451.5  &   5.9  \\
54687.89250   &     -417.5  &   7.9  \\
54713.82938   &     -432.2  &   7.4  \\
54729.77925   &     -409.6  &   8.3  \\
54744.74995   &     -383.3  &  11.7  \\
54759.92066   &     -403.3  &   6.5  \\
54774.68442   &     -402.6  &   8.0  \\
54786.85953   &     -411.2  &   7.6  \\
54794.61064   &     -361.8  &   6.8  \\
54812.59536   &     -377.5  &   7.1  \\
54821.55344   &     -378.8  &   7.0  \\
54857.66375   &     -375.1  &   9.3  \\
54883.60178   &     -336.9  &   6.0  \\
55048.92340   &     -397.6  &   4.9  \\
55073.84747   &   -407.6 &  6.0      \\
55102.77619   &   -417.5 &  3.4      \\
55172.57767   &   -357.1 &  8.3      \\
55198.71716   &   -352.9 &  4.9      \\
55222.66132   &   -343.9 &  7.1      \\
55246.58682   &   -300.8 &  7.5      \\
55246.61535   &   -308.4 &  8.5      \\
55249.61437   &   -328.3 &  6.2      \\
55444.83373   &   -387.4 &  4.5      \\
55468.74896   &   -368.4 &  7.1      \\
55493.71079   &   -406.5 &  5.5      \\
55493.71948   &   -404.3 &  6.3      \\
55498.68716   &   -347.8 &  5.1      \\
55513.63941   &   -366.4 &  6.4      \\
\enddata                                            
\end{deluxetable}

\begin{deluxetable}{lcc}
\tablecolumns{3}
\tablewidth{0pt}
\tablecaption{Relative RVs of HD 96127\label{table5}}
\tablehead{\colhead{Epoch [MJD]} & \colhead{RV [m s$^{-1}$]} & \colhead{$\sigma_{RV}$ [m s$^{-1}$]}}
\startdata
 53026.53749  &   -777.2  &      9.0   \\
 53026.54185  &   -775.3  &      8.5   \\
 53367.37012  &  -1114.1  &      4.5   \\
 53686.48016  &   -907.8  &      4.5   \\
 53721.40397  &   -804.9  &      5.9   \\
 53729.38407  &   -900.5  &      5.1   \\
 53736.34525  &   -916.9  &      7.8   \\
 53743.33658  &   -811.9  &      7.1   \\
 53752.29643  &   -879.8  &      6.4   \\
 53764.28676  &   -876.5  &      4.1   \\
 53794.19661  &   -875.5  &      6.9   \\
 53833.10419  &   -817.3  &      3.7   \\
 53844.28240  &   -832.5  &      5.4   \\
 53844.28410  &   -831.7  &      4.3   \\
 53844.28580  &   -826.9  &      6.1   \\
 53865.23660  &   -918.7  &      5.3   \\
 53865.23833  &   -920.6  &      5.4   \\
 53865.24019  &   -927.9  &      3.3   \\
 53865.24261  &   -919.8  &      2.6   \\
 53865.24509  &   -924.6  &      4.9   \\
 53877.19750  &   -905.2  &      7.7   \\
 53901.12750  &   -991.1  &      9.0   \\
 53901.12920  &   -992.5  &      6.9   \\
 53901.13090  &  -1003.2  &      7.8   \\
 54035.51530  &   -966.9  &     10.2   \\
 54035.51907  &   -985.7  &      7.3   \\
 54045.50512  &   -949.8  &      6.4   \\
 54080.41056  &  -1042.7  &      4.6   \\
 54121.29308  &   -937.0  &      5.4   \\
 54121.29877  &   -936.7  &      5.0   \\
 54156.19835  &   -870.1  &      8.3   \\
 54156.43557  &   -889.9  &      6.5   \\
 54190.11983  &   -838.9  &      4.6   \\
 54194.11568  &   -908.2  &      5.9   \\
 54194.11953  &   -897.9  &      5.9   \\
 54212.26824  &   -827.2  &      6.2   \\
 54212.27186  &   -843.2  &      6.1   \\
 54224.23445  &   -894.3  &      7.0   \\
 54242.19050  &   -943.1  &      5.1   \\
 54264.13327  &   -798.4  &      9.5   \\
 54437.44407   &  -867.4   &     5.0   \\
 54480.32430   &  -869.3   &    16.8   \\
 54483.30162   & -1021.7   &     5.2   \\
 54506.47125   &  -860.1   &     6.0   \\
 54527.19861   &  -997.1   &     9.3   \\
 54548.14498   &  -992.4   &     6.1   \\
 54553.12181   &  -930.6   &     6.5   \\
 54560.09780   &  -967.2   &     6.4   \\
 54597.23256   & -1018.5   &     7.1   \\
 54870.48390   &  -934.4   &     7.6   \\
\enddata                                          
\end{deluxetable}                                 

\begin{deluxetable}{lccccccc}
\tablecolumns{8}
\tabletypesize{\footnotesize}
\tablewidth{0pt}
\tablecaption{Giant Stars with Known Planets\label{table6}}
\tablehead{\colhead{Name} & \colhead{B-V} & \colhead{log($g$)}  & \colhead{[Fe/H]} & \colhead{$M_{\star}/M_{\odot}$} & \colhead{$R_{\star}/R_{\odot}$}& \colhead{rms [m s$^{-1}$]} & \colhead{Reference}}
\startdata
BD +20 2457 & 1.25 & 3.17 & -1.0 & 2.8 & 49 & 60 & \citet{nie09b} \\
BD +48 738 & 1.25 & 2.24 & -0.2 & 0.74 & 11& 16 & This paper\\
HD 11977 & 0.93 & 2.90 & -0.21 & 1.91 & 10& 29 & \citet{set05}\\
HD 13189 & 1.47 & - & - & 4 & 50 & 42 & \citet{hat05}\\
HD 17092 & 1.25 & 3.0 & 0.18 & 2.3 & 11 & 16 & \citet{nie07}\\
HD 32518 & 1.11 & 2.10 & -0.15 & 1.13 & 10 & 18 & \citet{dol09}\\
HD 81688 & 0.99 & 2.22 & -0.36 & 2.1 & 13 & 24 & \citet{sat08a}\\
HD 96127 & 1.50 & 2.06 & -0.24 & 0.91 & 20 & 50 & This paper\\
HD 102272 & 1.02 & 3.07 & -0.26 & 1.9 & 10 & 15 & \citet{nie09a}\\
HD 104985 & 1.03 & 2.62 & -0.35 & 2.3 & 11 & 27 & \citet{sat03}\\
HD 119445 & 0.88 & 2.40 & 0.04 & 3.9 & 20 & 13.7 & \citet{omi09}\\
HD 139357 & 1.20 & 2.9 & -0.13 & 1.35 & 12 & 14 & \citet{dol09}\\
HD 145457 & 1.04 & 2.77 & -0.14 & 1.9 & 10 & 10 & \citet{sat10}\\
HD 173416 & 1.04 & 2.48 & -0.22 & 2 & 13 & 18.5 & \citet{liu09}\\
HD 180314 & 1.00 & 2.98 & 0.20 & 2.6 & 9 & 13 & \citet{sat10}\\
HD 240210 & 1.65 & 1.55 & -0.18 & 1.25 & 11 & 25 & \citet{nie09b}\\
HD 240237 & 1.68 & 1.68 & -0.26 & 1.69 & 20 & 34 & This Paper\\
NGC 2423 No 3 & 1.21& - & 0.14 & 2.4 & - & 18 & \citet{lov07})\\
NGC 4349 No 127 & 1.46& - & -0.12 & 3.9 & - & 13 & \citet{lov07}\\
beta Gem & 1.00 &2.68 & 0.19 & 1.7 & 9 & - & \citet{hat93}\\
gamma Leo A & 1.13 & 1.59 & -0.49 & 1.23 & 40 & 43 & \citet{han10}\\ 
epsilon Tau &1.01 & 2.64 & 0.17 & 2.7 & 14 & 10 & \citet{sat07}\\
iota Dra & 1.18 & 2.24 & 0.03 & 1.05 & 13 & 10 & \citet{fri02}\\
xi Aql & 1.02 & 2.66 & -0.21 & 2.2 & 12 & 22 & \citet{sat08a}\\
4 UMa & 1.20& 1.8 & -0.25 & 1.23 & - & 30 & \citet{dol07}\\
11 Com & 0.99 & 2.5 & -0.35 & 2.7 & 19 & 25.5 & \citet{liu08}\\
11 UMi & 1.40 & 1.60 & 0.04 & 1.8 & 24 & 28 & \citet{dol09}\\
14 And & 1.03 & 2.63 & -0.24 & 2.2 & 11 & 20 & \citet{sat08b}\\
18 Del &0.93 & 2.82 & -0.05 & 2.3 & 9 & 15 & \citet{sat08a}\\
42 Dra & 1.19 & 1.71 & -0.46 & 0.98 & 22 & 14 & \citet{dol09}\\
81 Cet & 1.02 & 2.35 & -0.06 & 2.4 & 11 & 9 & \citet{sat08b}\\
\enddata                                          
\end{deluxetable}


\clearpage

\begin{thebibliography}

\bibitem[Alonso et al.(1999)]{alo99} Alonso, A., Arribas, S. \& C. Mart\'inez-Roger, 1999, A\&ASS, 140, 261 
\bibitem[Alonso et al.(2000)]{alo00} Alonso, A., Salaris, M., Arribas, S., Martinez-Roger, S., \& Asensio Ramos, A. 2000, A\&A, 355, 1060
\bibitem[Barban et al.(2004)]{bar04} Barban, C., et al. 2004, in Proceedings of the SOHO 14 / GONG 2004 Workshop, 559, 113 
\bibitem[Baudin et al.(2011)]{bau11} Baudin, F., et al. 2011, A\&A, 529, 84 
\bibitem[Benedict et al.(2010)]{ben10} Benedict, G. F., et al. 2010, AJ, 139, 1844
\bibitem[Benz \& Mayor(1984)]{ben84} Benz, W., \& Mayor, M.\ 1984, \aap, 138, 183 
\bibitem[Bowler et al.(2010)]{bow10} Bowler, B. P., Johnson, J. A., Marcy, G. W., Henry, G. W., Peek, K. M. G., Fischer, D. A., Clubb, K. I. et al., 2010, ApJ, 709, 396
\bibitem[Butler et al.(1996)]{but96} Butler, R. P., Marcy, G. W., Williams, E., McCarthy, C. \& Dosanjh, P., 1996, PASP, 108, 500
\bibitem[Butler et al.(2006)]{but06} Butler, R. P., et al. 2006, ApJ, 646, 505 
\bibitem[Chambers(1999)]{cha99} Chambers, J. E. 1999, MNRAS, 304, 793
\bibitem[Ciardi et al.(2011)]{cia11} Ciardi, D. R., et al. 2011, AJ, 141, 108
\bibitem[de Ridder et al.(2009)]{der09} de Ridder, J., et al. 2009, Nature, 459, 398
\bibitem[D\"ollinger et al.(2007)]{dol07} D\"ollinger, M. P., Hatzes, A. P., Pasquini, L., Guenther, E. W., Hartmann, M., Girardi, L. \& Esposito, M., 2007, A\&A, 472, 649
\bibitem[D\"ollinger et al.(2009)]{dol09} D\"ollinger, M. P., Hatzes, A. P., Pasquini, L., Guenther, E. W., Hartmann, M. \& Girardi, L., 2009, A\&A, 499, 935
\bibitem[D\"ollinger et al.(2011)]{dol11} D\"ollinger, M. P., et al., 2011, in AIP Conference Proc. 1331, Planetary Systems Beyond the Main Sequence, 79
\bibitem[Fischer \& Valenti(2005)]{fis05} Fischer, D. A. \& Valenti, J., 2005, ApJ, 622, 1102
\bibitem[Frink et al.(2002)]{fri02} Frink, S., Mitchell, D. S., Quirrenbach, A., Fischer, D. A., Marcy, G. W. \& Butler, R. P., 2002, ApJ, 576, 478
\bibitem[Ghezzi et al.(2010)]{ghe10} Ghezzi, L., Cunha, K., Schuler, S. C. \& Smith, V. V., 2010, ApJ, 725, 721
\bibitem[Gilliland et al.(2011)]{gil11} Gilliland, R. L. et al., ApJ, 726, 2
\bibitem[Girardi et al.(2000)]{gir00} Girardi, L., Bressan, B., Bertelli, G., Chiosi, C. 2000, A\&AS, 141, 371
\bibitem[Halbwachs et al.(2005)]{hal05} Halbwachs, J. L., Mayor, M. \& Udry, S. 2005, A\&A, 431, 1129
\bibitem[Han et al.(2010)]{han10} Han, I., Lee, B. C., Kim, K. M., Mkrtichian, D. E., Hatzes, A. P. \& Valyavin, G., 2010, A\&A, 509, A24 
\bibitem[Hatzes et al.(1993)]{hat93} Hatzes, A. P. \& Cochran, W. D., 1993, ApJ, 413, 339 
\bibitem[Hatzes \& Cochran(1998)]{hat98} Hatzes, A. P. \& Cochran, W. D., 1998, in ASP Conf. Ser. 154, The Tenth Cambridge Workshop on Cool Stars, Stellar Systems and the Sun, Edited by R. A. Donahue and J. A. Bookbinder, 311
\bibitem[Hatzes(2002)]{hat02} Hatzes, A. P. 2002, Astron. Nachr., 323, 392
\bibitem[Hatzes et al.(2005)]{hat05} Hatzes, A. P., Guenther, E. W., Endl, M., Cochran, W. D., D\"ollinger, M. P., \& Bedalov, A. 2005, A\&A, 437, 743
\bibitem[Hatzes et al.(2006)]{hat06} Hatzes et al., 2006, A\&A, 457, 335
\bibitem[Hatzes et al.(2007)]{hat07} Hatzes, A. P., D\"olinger, M. P. \& Endl, M., 2007, CoAst, 150, 115
\bibitem[Hatzes \& Zechmeister(2008)]{hat08} Hatzes, A. P. \& Zechmeister, M., 2008, in Journal of Physics: Conference Series 118, 12016
\bibitem[Hekker et al.(2006)]{hek06} Hekker, S., Reffert, S., Quirrenbach, A., Mitchell, D. S., Fischer, D. A., Marcy, G. W. \& Butler, R. P., 2006, A\&A, 454, 943
\bibitem[Hekker \& Mel\'endez(2007)]{hek07} Hekker, S. \& Mel\'endez, J., 2007, A\&A, 475, 1003
\bibitem[Hekker et al.(2008)]{hek08} Hekker, S., Snellen, I. A. G., Aerts, C., Quirrenbach, A., Reffert, S. \& Mitchell, D. S., 2008, A\&A, 480, 215
\bibitem[Hekker(2010)]{hek10} Hekker, S., 2010, AN, 331, 1004
\bibitem[Hekker et al.(2011)]{hek11} Hekker et al., 2011, arXiv:1103.0141v1, to be published in MNRAS
\bibitem[H$\o$g et al.(2000)]{hog00} H$\o$g, E., et al., 2000, A\&A, 355, L27
\bibitem[Ida \& Lin(2004)]{ida04} Ida, S. \& Lin, D. N. C., 2004, ApJ, 616, 567
\bibitem[Johnson et al.(2007)]{joh07} Johnson, J. A. et al., 2007, ApJ, 665, 785
\bibitem[Johnson et al.(2010)]{joh10} Johnson, J., Aller, K. M., Howard, A. W \& Crepp, J. R., 2010, ApJ, 122, 905
\bibitem[Kjeldsen \& Bedding(1995)]{kje95} Kjeldsen, H., \& Bedding, T. R. 1995, A\&A, 293, 8
\bibitem[Kozai(1962)]{koz62} Kozai, Y.. 1962, AJ, 67, 591
\bibitem[Kraus et al.(2011)]{kra11} Kraus, A.L., Ireland, M.J., Martinache, F., Hillenbrand, L.A., 2011, ApJ, 731, 8
\bibitem[Liu et al.(2008)]{liu08} Liu, Y.-J., et al., 2008, ApJ, 672, 553
\bibitem[Liu et al.(2009)]{liu09} Liu, Y.-J., Sato, B., Zhao, G. \& Ando, H., 2009, RAA, 9, 1
\bibitem[Lovis \& Mayor(2007)]{lov07} Lovis, C., \& Mayor, M. 2007, A\&A, 472, 657
\bibitem[Mart\'inez Fiorenzano(2005)]{mar05} Mart\'inez Fiorenzano, A. F., et al. 2005, A\&A, 442, 775
\bibitem[Niedzielski et al.(2007)]{nie07} Niedzielski, A., et al., 2007, ApJ, 669, 1354
\bibitem[Niedzielski \& Wolszczan(2008)]{nie08} Niedzielski, A. \& Wolszczan, A., 2008, ASPCS, 398, 71
\bibitem[Niedzielski et al.(2009a)]{nie09a}  Niedzielski, A., Go\'zdziewski, A.,  Wolszczan, A., Konacki, M., Nowak, G. \& Zieli\'nski, P., 2009, ApJ, 693, 276
\bibitem[Niedzielski et al.(2009b)]{nie09b} Niedzielski, A., Nowak, G., Adam\'ow, M. \& Wolszczan, A., 2009, ApJ, 707, 768
\bibitem[Nordhaus et al.(2010)]{nor10} Nordhaus, J., Spiegel, D. S., Ibgui, L., Goodman, J., Burrows, A., 2010, MNRAS, 408, 631
\bibitem[Nowak et al.(2010)]{now10} Nowak, G., Niedzielski, A. \& Wolszczan, A., 2010, in EAS Publications Series 42,
Extra-solar Planets in Multi-body Systems. Theory and Observations, ed.
K. Go\'zdziewski, A. Niedzielski, \& J. Schneider, 165
\bibitem[Nowak et al.(2011)]{now11} Nowak, G. et al. 2011, in prep.
\bibitem[Omiya et al.(2009)]{omi09} Omiya, M., et al., 2009, PASJ, 61, 825
\bibitem[Omiya et al.(2011)]{omi11} Omiya, M., et al., 2011, in AIP Conference Proc. 1331, Planetary Systems Beyond the Main Sequence, 122
\bibitem[Oppenheimer \& Hinkley(2009)]{opp09} Oppenheimer, B.R., Hinkley, S., 2009, ARA\&A, 47, 253
\bibitem[Pasquini et al.(2007)]{pas07} Pasquini, L. et al., 2007, A\&A, 473, 979
\bibitem[Pasquini et al.(2008)]{pas08} Pasquini,  L., D\"ollinger, M. P., Hatzes, A., Setiawan, J., Girardi, L., da Silva, L., de Medeiros, J. R., \& Weiss, A.
, 2008, in IAU Symp. 249, Exoplanets: Detection, Formation \& Dynamics, ed. Y.-S. Sun, S. Ferraz-Mello, \& J.-L. Zhou (Cambridge: Cambridge Univ. Press), 209
\bibitem[Perryman \& ESA(1997)]{per97} Perryman, M. A. C., \& ESA 1997, in ESA Special Publication 1200, The Hipparcos and Tycho Catalogues. Astrometric and Photometric Star Catalogues derived from the ESA Hipparcos Space Astrometry Mission
(Noordwijk: ESA)
\bibitem[Pollacco et al.(2006)]{pol06} Pollacco, D. L., et al., 2006, PASP, 118, 1407
\bibitem[Press et al.(1992)]{pre92} Press, W. H., Teukolsky, S. A., Vetterling, W. T., \& Flannery, B. P. 1992, Numerical Recipes in FORTRAN. The Art of Scientific Computing (2nd ed.; Cambridge: Cambridge Univ. Press)
\bibitem[Queloz et al.(2001)]{que01} Queloz, D., Henry, G. W., Sivan, J. P., Baliunas, S. L., Beuzit, J. L., Donahue, R. A., Mayor, M. et al., 2001,A\&A, 379, 279
\bibitem[Quirrenbach et al.(2011)]{qui11} Quirrenbach, A., Reffert, S. \& Bergmann, C., 2011, in AIP Conference Proc. 1331, Planetary Systems Beyond the Main Sequence, 102
\bibitem[Ramsey et al.(1998)]{ram98} Ramsey, L. W., 1998, Proc. SPIE, 3352, 34
\bibitem[Salasnich et al.(2000)]{sal00} Salasnich, B., Girardi, L., Weiss, A. \& Chiosi, C., 2000, 361, 1023
\bibitem[Sato et al.(2003)]{sat03} Sato, B., et al., 2003, ApJ, 597, 157
\bibitem[Sato et al.(2007)]{sat07} Sato, B., et al., 2007, ApJ, 661, 527
\bibitem[Sato et al.(2008a)]{sat08a} Sato, B., et al., 2008, PASJ, 60, 539
\bibitem[Sato et al.(2008b)]{sat08b} Sato, B., et al., 2008, PASJ, 60, 1317
\bibitem[Sato et al.(2010)]{sat10} Sato, B., et al., 2010, PASJ, 62, 1063
\bibitem[Setiawan et al.(2005)]{set05} Setiawan, J., Rodmann, J., da Silva, L., Hatzes, A., Pasquini,  L., von der L\"uhe, O., de Medeiros, J. R., D\"ollinger, M. P. \& Girardi, L., 2005, A\&A, 437, L31
\bibitem[Shetrone et al.(2007)]{she07} Shetrone, M., et al. 2007, PASP, 119, 556
\bibitem[Straizys \& Kuriliene(1981)]{str81} Straizys, V. \& Kuriliene, G., 1981, Ap\&SS, 8
\bibitem[Stumpff(1980)]{stu80} Stumpff, P. 1980, A\&AS, 41, 1
\bibitem[Takeda et al.(2002)]{tak02} Takeda, Y., Sato, B., Kambe, E., Sadakane, K., \& Ohkubo, M., 2002, PASJ, 54, 104
\bibitem[Takeda et al.(2005a)]{tak05a} Takeda, Y., Ohkubo, M., Sato, B., Kambe, E., \& Sadakane, K., 2005, PASJ, 57, 27
\bibitem[Takeda et al.(2005b)]{tak05b} Takeda, Y., Sato, B., Kambe, E., Izumiura, H., Masuda, S., \& Ando, H., 2005, PASJ, 57,109
\bibitem[Tull(1998)]{tul98} Tull, R. G., 1998, Proc. SPIE, 3355, 387
\bibitem[Villaver \& Livio(2009)]{vil09} Villaver E. \& Livio, M., 2009, ApJ, 705, L81
\bibitem[Winn et al.(2010)]{win10} Winn, J. N., et al., 2010, ApJ, 718, 575
\bibitem[Wo{\'z}niak et al.(2004)]{woz04} Wo{\'z}niak, P.~R., et al.\ 2004, \aj, 127, 2436
\bibitem[Wright et al.(2007)]{wri07} Wright, J. T., et al. 2007, ApJ, 657, 533
\bibitem[Zechmeister et al.(2008)]{zec08} Zechmeister, M., Reffert, S., Hatzes, A. P., Endl, M. \& Quirrenbach, A., 2008, A\&A, 491, 531
\bibitem[Zieli\'nski et al.(2011 in prep.)]{zie11} Zieli\'nski, P., Niedzielski, A., Adam\'ow, M. \& Wolszczan, A., 2010, EAS Publications Series, 42, 201

\end{thebibliography}
\end{document}